\documentclass[prb,aps,onecolumn]{revtex4}
\input epsf
\input pstricks
\begin{document}
\def\etr{\varepsilon_{\rm tr}}
\def\beq{\begin{equation}}
\def\eeq{\end{equation}}
\def\lmin{L_{\rm min}}
\def\lmax{L_{\rm max}}
\def\moy#1{\langle #1\rangle}
\title{Statistical evaporation of rotating clusters.
II. Angular momentum distribution}
\author{P. Parneix}
\affiliation{Laboratoire de Photophysique Mol\'eculaire, B\^at. 210,
Universit\'e Paris-Sud, F91405 Orsay Cedex, France}
\author{F. Calvo}
\affiliation{Laboratoire de Physique Quantique, IRSAMC, Universit\'e Paul
Sabatier, 118 Route de Narbonne, F31062 Toulouse Cedex, France}
\begin{abstract}
The change in the angular momentum of an atomic cluster following
evaporation is investigated using rigorous phase space theory and
molecular dynamics simulations, with an aim at the possible rotational
cooling and heating effects. Influences of the shape of the
interaction potential, anharmonicity of the vibrational density of
states (DOS), and the initial distribution of excitation energies are
systematically studied on the example of the Lennard-Jones cluster
LJ$_{14}$. For this system, the predictions of PST are in quantitative
agreement with the results of the simulations, provided that the correct
forms for the vibrational density of states and the interaction
potential are used. The harmonic approximation to the DOS is used to
obtain explicit forms for the angular momentum distribution in larger
clusters. These are seen to undergo preferential cooling when thermally
excited, and preferential heating when subject to a strong vibrational
excitation.
\end{abstract}
\maketitle

\section{Introduction}

When a polyatomic system is vibrationally excited, it may undergo
spontaneous dissociation on a suitable time scale. Most often,
different dissociation channels are available, which may or may not be
selected upon energetic criteria only. Rotational excitations are also
very important, because they directly influence the relative amount of
kinetic energy released in translational modes. Kinetic energies released
are of constant use in experiments for extracting information about
binding energies, temperatures, and more generally nanocalorimetry
data (see Refs. \onlinecite{brech01,vogel01,brech02} and references therein).

Rotational excitations practically act in two distinct ways. The first is
direct, and reflects the greater ease of fragmenting the system due to
centrifugal forces. The second way is indirect, and lies in the
coupling between vibrational and rotational motion: because the system
is not rigid, it may rearrange so as to reduce its angular velocity,
thereby enhancing its stability. When an isolated cluster dissociates
into two fragments, the angular momentum $\vec J$ is splitted into two
parts, namely the orbital momentum $\vec L$ of the fragments in their
relative motion, and the remaining angular momentum $\vec J_r$
associated with the fragments themselves. It is very easy to see that,
at low initial value of $\vec J$, $\vec L$ nearly compensates $\vec
J_r$. Hence, for a large system emitting a single atom, rotational
heating is expected at low $J$. On the opposite, rapidly rotating
clusters are more likely to lose angular momentum following
evaporation, a phenomenon known as rotational cooling. These aspects
of fragmentation have been previously investigated by
Stace,\cite{stace91} who used simple statistical models to emphasize
the role of rotational temperature in the interpretation of experiments.

In our previous work,\cite{calvo_2003} we performed molecular
dynamics (MD) simulations as well as theoretical calculations based on
the phase space theory (PST) to extract the distributions of kinetic
energies released upon evaporation in small Lennard-Jones (LJ) atomic
clusters. Our main result was to show that PST could be quantitatively
accurate in a broad range of total energies and angular momenta,
provided that the vibrational and rotational densities of states, as
well as the interaction potential between the fragments, were all
correctly accounted for. This result was previously proved by
Weerasinghe and Amar for nonrotating clusters,\cite{wa} and partially
observed for aluminium clusters by Peslherbe and Hase,\cite{ph0,ph} though
these authors used a simple $C/r^n$ form for the dissociation
potential, which induced some deviations between theory and simulation
at large excitations.

Here we will focus on angular momentum properties, namely their
distributions and mean values, and how they are affected by the
initial excitation. The present approach follows our previous
effort,\cite{calvo_2003} with the same aim at assessing phase space
theory in a quantitative comparison with simulations. In the next
section, we briefly recall the main equations underlying the PST
calculation of the angular momentum properties of the product cluster,
and we treat separately the simple cases of an harmonic vibrational
density of states, an interaction potential with the form $C/r^n$,
or an initially nonrotating cluster. Our application to small and
medium size Lennard-Jones clusters for various types of excitations is
treated in Sec.~\ref{sec:res}. In Sec.~\ref{sec:ccl} we
discuss our results with respect to the works by Stace\cite{stace91}
and those by Peslherbe and Hase,\cite{ph} before finally concluding.

\section{Phase space theory}
\label{sec:pst}

In this Section, we give the main expressions for the angular momentum
distribution of the product cluster following evaporation.
To obtain such information, conservation of total angular momentum and
energy has to be included, which is the case in the phase space
theory developed by Pechukas and Light,\cite{pechukas65,light67}
Klots,\cite{klots71,klots76} and Chesnavich and Bowers.\cite{cb1,cb,cb2}

Here we consider a parent cluster characterized by a rotational angular
momentum $J$ and a total rovibrational energy $E$. We denote by $J_r$ the
rotational angular momentum of the product cluster after dissociation.
The unnormalized probability of finding a dissociation event characterized by
the angular momentum $J_r$ of the product cluster within $dJ_r$ and the energy
released $\etr$ within $d\etr$ can be written as\cite{cb}
\beq
{\cal P}(\etr,J_r;E,J) = \Omega_n^{(J_r)}(E-E_0-\etr)
\int_{\lmin(\etr,J_r)}^{\lmax(\etr,J_r)} \Gamma(\varepsilon_r^*,J_r) dL.
\label{eq:peej}
\eeq
In this equation, $\Omega_n^{(J_r)}$ is the vibrational density of
states of the product cluster having angular momentum $J_r$,
$\Gamma(\varepsilon_r^*,J_r)$ is the available volume of states with
rotational energy lower than $\varepsilon_r^*$ in angular momentum
space. For a given $\etr$, the rotational density of states is given
by the sum of this quantity over the accessible area in the $(J_r,L)$
plane, with $L$ the orbital angular momentum. This area is limited by
mechanical (angular momentum and energetic) constraints, and will be
discussed below. If both $\etr$ and $J_r$ are fixed, the integration
must be performed in a range of $L$ limited by $\lmin$ and $\lmax$. In
this paper, we are interested in the distribution of angular momentum
of the product cluster. The probability of finding a particular value
of $J_r$ within $dJ_r$ is simply the integral of ${\cal P}$ over all
possible values of $\etr$:
\beq
P(J_r;E,J) = \int_{\etr^{\rm min}}^{E-E_0} {\cal P}(\etr,J_r; E,J) d\etr
\label{eq:reej}
\eeq
In PST the products are located at the transition state. Here we will
consider rather large clusters, and our main assumption will be to treat the
evaporative system LJ$_{n+1}\to$LJ$_n+$LJ within the sphere$+$atom model. In
this case, the available volume $\Gamma(\varepsilon_r^*,J_r)$ is simply given
by $2J_r$.\cite{jarrold} Therefore the probability of finding a dissociation
event with angular momentum $J_r$, starting from a total energy $E$ and a total
angular momentum $J$ is given by:
\beq P(J_r;E,J) = P_0\times  2J_r \int_{\etr^{\rm min}}^{E-E_0}
\Omega_n^{(J_r)}(E-E_0-\etr) [\lmax(\etr,J_r) - \lmin(\etr,J_r)] d\etr
\label{eq:reejneww},
\eeq
where the normalization constant $P_0$ accounts for channel and
rotational degeneracies, and also depends on the parent density of
states $\Omega_{n+1}^{(J)}(E)$. In the following, we will make use of
the notation $\Delta L(\etr,J_r)=\lmax(\etr,J_r)-\lmin(\etr,J_r)$. The
problem is now reduced to determining $\Delta L(\etr,J_r)$.

\begin{figure}[htb]
\setlength{\epsfxsize}{8cm} \centerline{\epsffile{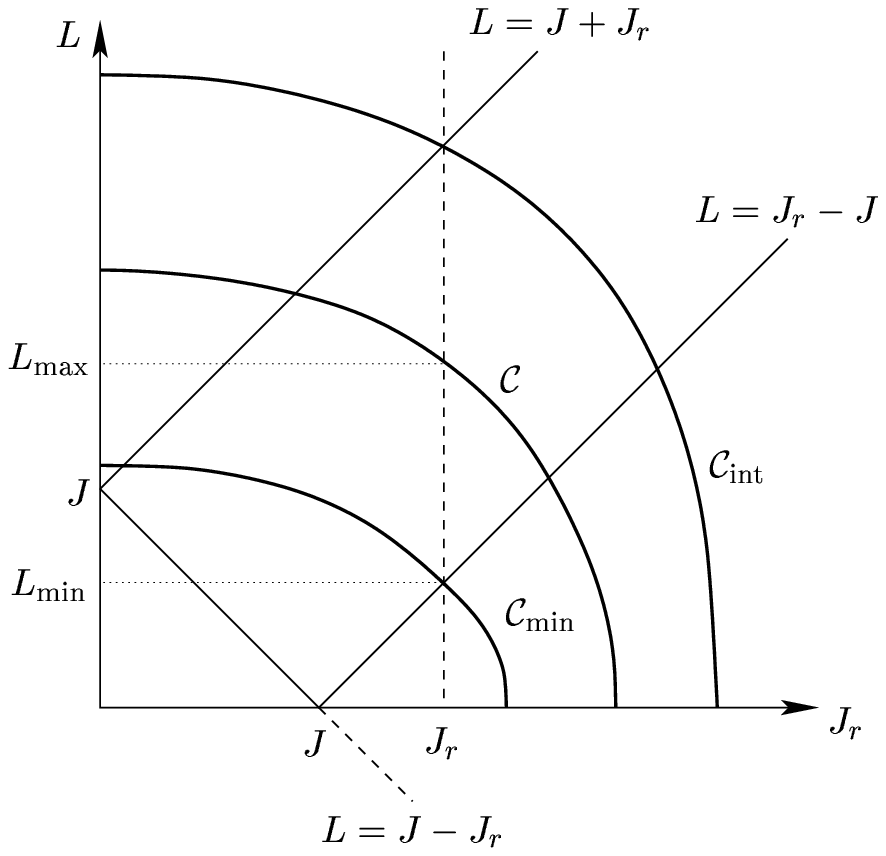}}
\caption{Schematic representation of the $(J_r,L)$ integration plane.}
\label{fig:schema}
\end{figure}

We first consider the case of a radial dissociation potential $V(r)$ given by
$V(r)=-C_n/r^p$, with $p>2$. Since the kinetic energy must be positive
at the centrifugal barrier, the well known
energetic constraint can be easily found:
\beq B_nJ_r^2 + L^{2p/(p-2)}/\Lambda_p \leq \etr, \label{eq:cond7} \eeq
where we have introduced the rotational constant $B_n$ of the product
cluster LJ$_n$, and the Langevin parameter $\Lambda_p$ given as a
function of $p$, $C_n$, $\hbar$ and the reduced mass $\mu_n$ by
\beq
\Lambda_p= \frac{2}{p-2}C_n^{2/(p-2)}\left( \frac{\mu_n
p}{\hbar^2}\right)^{p/(p-2)}.
\label{eq:lambda}
\eeq
The second boundary comes from the conservation of angular momentum, ${\vec J}
= {\vec J}_r + {\vec L}$, or more conveniently:
\beq |J_r - L| \leq J \leq J_r+L.
\label{eq:cond8} \eeq
Let us denote by ${\cal C}$ the set of $(J_r,L)$ points that fulfill
the equation $\etr=L^{2p/(p-2)}/\Lambda_p +B_nJ_r^2$. For a given
value of $J_r$, we define $\etr^{\rm min}$ (resp. $\etr^{\rm int}$) as
the values of $\etr$, which correspond to the intersection between
${\cal C}$ and $L=|J-J_r|$ (resp. $L=J+J_r$). We find
\beq \etr^{\rm min} = B_nJ_r^2+ \frac{|J-J_r|^{2p/(p-2)}}{\Lambda_p}
\mbox{~~~and~~~} \etr^{\rm int} = B_nJ_r^2+
\frac{(J+J_r)^{2p/(p-2)}}{\Lambda_p}
\label{eq:jrl} \eeq
Obviously, $\etr^{\rm min}$ and $\etr^{\rm int}$ can not be larger than the
maximum available energy $E-E_0$, and the largest value for $J_r$ will
be reached when $\etr^{\rm min}$=$E-E_0$. In Fig.~\ref{fig:schema}, a
simple scheme summarizes the integration procedure in the $(J_r,L)$
plane. In this figure, ${\cal C}_{\rm min}$ and ${\cal C}_{\rm int}$
are the contours ${\cal C}$ for $\etr=\etr^{\rm min}$ and
$\etr=\etr^{\rm int}$, respectively. When $\etr^{\rm min} \leq \etr
\leq \etr^{\rm int}$, $\Delta L(\etr,J_r)$ is given by ${\cal
L}(\etr,J_r) - |J-J_r|$, where ${\cal L}(\etr,J_r)=[\Lambda_p(\etr -
B_nJ_r^2)]^{(p-2)/2p}$. On the other hand, when $\etr^{\rm int} \leq
\etr \leq E-E_0$, $\Delta L(\etr,J_r) = (J+J_r) - |J-J_r|$.

One interesting case is the harmonic limit of this model. For this we
consider the harmonic VDOS for the product cluster,
$\Omega_n(E)\propto E^{s-1}$ with $s=3n-6$. After replacing the
expressions for $\Delta L(\etr,J_r)$ obtained above, equation
(\ref{eq:reejneww}) now becomes
\begin{eqnarray}
P_h(J_r;E,J)  &\propto& 2J_r \int_{\etr^{\rm min}}^{\etr^{\rm int}}
(E-E_0-\etr)^{s-1} {\cal L}(\etr,J_r) d\etr \nonumber \\
&-& 2J_r |J-J_r| \ \int_{\etr^{\rm min}}^{\etr^{\rm
int}} (E-E_0-\etr)^{s-1} d\etr \nonumber \\
&+& 2J_r \ (J+J_r - |J-J_r|) \int_{\etr^{\rm int}}^{E-E_0}
(E-E_0-\etr)^{s-1} d\etr,
\label{eq:peejr}
\end{eqnarray}
By considering the analytical expression for ${\cal L}(\etr,J_r)$
given previously and using the notation $\gamma=(p-2)/2p$, we thus obtain
\begin{eqnarray}
P_h(J_r;E,J) & \propto & 2J_r \Lambda_p^\gamma \int_{\etr^{\rm
min}}^{\etr^{\rm int}}
(E-E_0-\etr)^{s-1} (\etr-B_nJ_r^2)^\gamma d\etr \nonumber \\
&-&\frac{2J_r |J-J_r|}{s} [(E-E_0-\etr)^s]_{\etr^{\rm
int}}^{\etr^{\rm min}} \nonumber \\
&+& \frac{2J_r (J+J_r - |J-J_r|)}{s} [(E-E_0-\etr)^s]_{E-E_0}^{\etr^{\rm int}}.
\label{eq:peejr_inter}
\end{eqnarray}
This equation can be further simplified into
\begin{eqnarray}
P_h(J_r;E,J) &\propto& \frac{2J_r(J+J_r)}{s} (E-E_0-\etr^{\rm int})^s \nonumber
\\ &-& \frac{2J_r |J-J_r|}{s} (E-E_0-\etr^{\rm min})^s + 2J_r
\ \Lambda_p^\gamma {\cal I}_{s,\gamma} , \label{eq:peejr_final}
\end{eqnarray}
where ${\cal I}_{s,\gamma}$ is short for the integral in
Eq.~(\ref{eq:peejr_inter}).
This integral can be written explicitely as
\beq {\cal I}_{s,\gamma}=
\sum_{k=0}^{s-1} {s-1 \choose k} \frac{(-1)^{s-1-k}}{s-k+\gamma}
\frac{(E-E_0-B_nJ_r^2)^k}{\Lambda_p^{s-k+\gamma}}
\left[ (J+J_r)^{1+(s-k)/\gamma} - |J-J_r|^{1+(s-k)/\gamma} \right].
\label{eq:Int}
\eeq
All the formalism above has been derived by assuming an interaction
potential with the form $V(r)=-C_n/r^p$. As discussed in our previous
work,\cite{calvo_2003} this expression does not give a very good account of the
finite extension of the cluster, and a better representation of the
atom-cluster interaction is provided by $V(r)= -C_n/(r-r_0)^p$, with $r_0>0$. In
this case, and more generally for an arbitrary form of $V(r)$, the computation
of $\lmin(\etr,J_r)$ and $\lmax(\etr,J_r)$ must be carried out
numerically. For a series of $L$, the location $r^*(L)$ and the height
$\varepsilon^\dagger(L)$ of the centrifugal barrier is obtained. At a given
$\etr$, the integration boundaries become
\beq
\left\{ \begin{array}{l}
\varepsilon^\dagger(L)+B_nJ_r^2 \leq \etr, \\
|J_r-L|\leq J\leq J_r+L \end{array} \right.
\label{eq:snum}
\eeq
and the limits $\lmin(\etr,J_r)$ and $\lmax(\etr,J_r)$ can thus be
directly calculated from the knowledge of $\varepsilon^\dagger(L)$.

When we do not specify the harmonic limit to extract $P(J_r;E,J)$, equation
(\ref{eq:reejneww}) is used with the anharmonic vibrational densities of states
$\Omega^{(J_r)}(E)$, which includes some dependency with $J_r$.
To calculate $\Omega^{(J_r)}(E)$, we have used the Monte Carlo method proposed
in Ref.~\onlinecite{mcrot}, further improved with the parallel tempering
accelerating scheme.\cite{ptmc} The multiple histogram method\cite{histo} was
then used to estimate the configurational densities of states and, through a
simple convolution product, the total VDOS. In practice, we have neglected the
variations of $\Omega$ with $J_r$, and used the approximation $J_r\approx J$.
As will be seen below, this does not entail a significant error.

The dissociation potential $V(r)$ felt by an atom leaving the $n$-atom LJ cluster
was taken from our previous work,\cite{calvo_2003} and approximated by the form
$-C_6/(r-r_0)^6$ form. The coefficients $C_6$ and $r_0$ have been deduced from
a fitting of constrained Monte Carlo simulations at finite temperature, as
well as some Wang-Landau simulations. Their values are given in Ref.
\onlinecite{calvo_2003}.

For each value of angular momentum, we have also carried out one or
several sets of 5000
independent molecular dynamics trajectories, used as reference data.
The angular momentum distribution has been calculated
for different values of the initial total angular momentum between $J=0$ and $J
\approx 5$ LJ units.\cite{conversion} For the details of the MD simulations, we
also refer the reader to Ref.~\onlinecite{calvo_2003}.

\section{Results}
\label{sec:res}

The energetics of unimolecular evaporation has been considered in
our previous article,\cite{calvo_2003} and we only focus here on the
angular momentum distribution following the evaporation of a rotating atomic
cluster. In the form detailed above, phase space theory directly gives us
access to this distribution. The sharing between rotational energy of the
product cluster and the translational kinetic energy released can then be
analysed. 

\begin{figure}[htb]
\vbox to 8cm{
\includegraphics{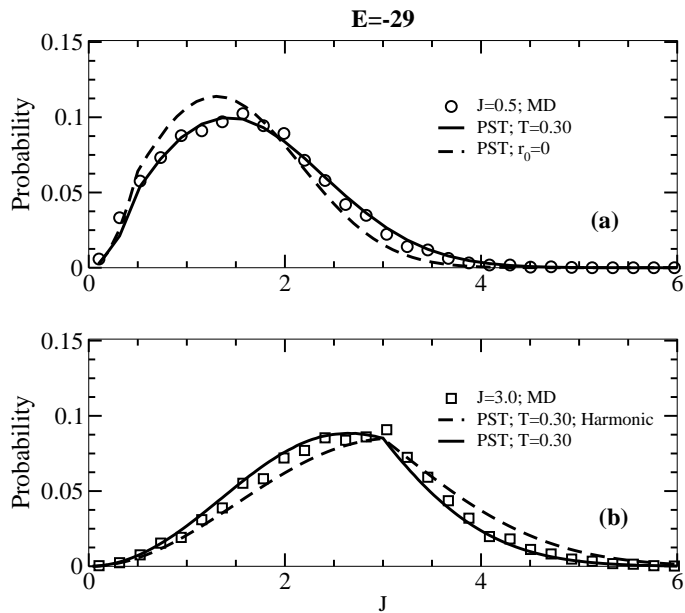}
\vfill}
\caption{Angular momentum distribution of $J_r$ for the
reaction LJ$_{14}\longrightarrow$LJ$_{13}+$LJ at $E=-29$, for two
different values of the initial angular momentum $J$. (a) $J=0.5$; (b) $J=3$.}
\label{fig:distriLJ13}
\end{figure}

We first  consider the unimolecular dissociation  involving the nearly
spherical LJ$_{13}$ product cluster. The reaction studied here is thus
LJ$_{14}\to$LJ$_{13}+$LJ. In Fig.~\ref{fig:distriLJ13} we have plotted
the angular momentum distribution of the LJ$_{13}$ product cluster, as
estimated from PST. The rotational constant $B$ of LJ$_{13}$ was taken
at $T=0.3$, such that $B=0.0444$ LJ units.\cite{calvo_2003} Two values
of  the  initial angular  momentum,  namely  $J=0.5$  and $J=3$,  were
considered  and the initial  rovibrational energy  was taken  equal to
$E=-29$ LJ units in both cases. Different approximations were assessed
and  a  good  overall  agreement  with MD  results  is  obtained.   In
Fig.~\ref{fig:distriLJ13}(a),  two  different  interaction  potentials
were selected  along with  the anharmonic density  of states  for both
calculations.  The  interaction potential corresponding  to $T=0.3$ is
computed from MC simulations,  and the $r_0=0$ approximation refers to
the long-range part of the exact LJ potential, with $C_6=4n$ units. As
can   be   seen   from   Fig.~\ref{fig:distriLJ13}(a),   the   $r_0=0$
approximation performs slightly less than the numerically exact radial
potential,  but still  allows  to reproduce  quite  accurately the  MD
results. The $r_0=0$ approximation  tends to underestimate the
angular momentum of the  product cluster, a feature previously noticed
by Peslherbe  and Hase.\cite{ph} In  Fig.~\ref{fig:distriLJ13}(b), the
harmonic  limit has been  tested for  $J=3$. Even  though a  very good
agreement  is  obtained  when  considering the  anharmonic  DOS,  some
discrepancy with MD  becomes to appear in the  harmonic case. A change
in slope  has to be noted at  $J_r=J$ in Figs.~\ref{fig:distriLJ13}(a)
and  especially  \ref{fig:distriLJ13}(b).
This  discontinuity  in   the  derivative  of
$P(J_r;E,J)$ is due to the modulus $|J_r-J|$ term in the definition of
$\etr^{\rm min}$. Taking the derivative of Eq.~\ref{eq:reejneww} as a
function of $J_r$ actually leads to the following value for the change
in slope accross $J=J_r$:
\begin{eqnarray}
\Delta \left( \frac{\partial P}{\partial J_r}\right) &=&
\left. \frac{\partial P}{\partial J_r}\right|_{J=J_r^+} -
\left. \frac{\partial P}{\partial J_r}\right|_{J=J_r^-} \nonumber \\
&=& -4P_0 J\int_{B_nJ^2}^{E-E_0} \Omega_n^{(J)}(E-E_0-\etr)d\etr.
\label{eq:discontinuity}
\end{eqnarray}
Therefore the slope necessary decreases at $J_r=J$, independently of
the dissociation potential. In order to put this effect into evidence,
we have run 15 sets of 5000 trajectories at $J=3.0$
[see Fig.~\ref{fig:distriLJ13}(b)].
Within this statistics, a change in the slope around $J_r=J$ can be
effectively characterized, as seen in the statistical PST formalism.

\begin{figure}[htb]
\vbox to 8cm{
\includegraphics{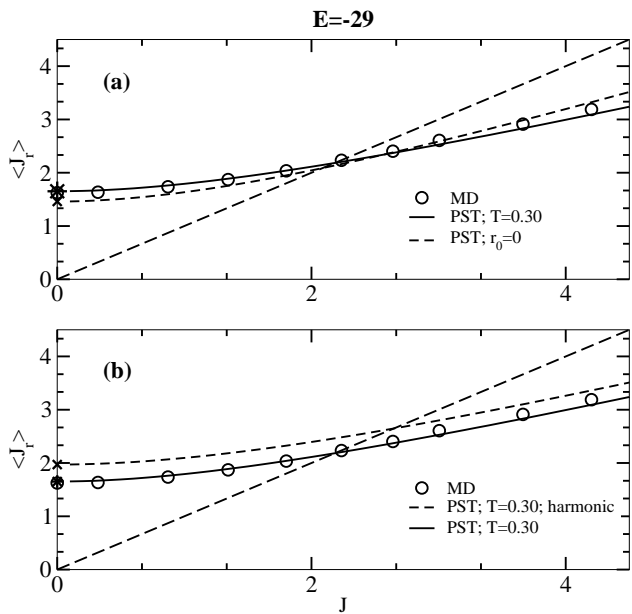}
\vfill}
\caption{$\moy{J_r}$ versus $J$ for the reaction
LJ$_{14}\longrightarrow$LJ$_{13}+$LJ at $E=-29$ within different 
approximations: (a) effect of the radial potential; (b) effect of the
anharmonicity of the PES. The $J=0$ values, obtained from
Eq.~(\ref{eq:J=0}), are represented by stars or crosses.}
\label{fig:JmeanLJ13}
\end{figure}

In Fig.~\ref{fig:JmeanLJ13}, the mean  angular momentum of the product
cluster $\moy{J_r}$ has  been plotted as a function of  $J$ at the
rovibrational energy $E=-29$. The  same assumptions as in the previous
figure have been tested. The line $\moy{J_r}=J$ has also been shown in
order  to visualize  more  easily the  rotational  cooling or  heating
effects.  First of  all it is striking that  the statistical PST model
allows to perfectly reproduce the  MD results when both the anharmonic
DOS and  the numerically exact  radial potential are  considered.  The
$r_0=0$ approximation  tends to  underestimate $\moy{J_r}$ in  the low
$J$  regime.  As  $J$  goes to  0,  one recovers  the  exact limit  of
Eq.~(\ref{eq:reejneww}) given by
\beq    P(J_r;E,J=0)    =     P_0    \times    2J_r    \int_{B_nJ_r^2+
\varepsilon^\dagger(J_r)}^{E-E_0}    \Omega_n^{(J_r)}(E-E_0-\etr)d\etr.
\label{eq:J=0}
\eeq
The harmonic approximation to this equation is straightforward:
\beq   P(J_r;E,J=0)    \propto   J_r   \left[    E-E_0   -B_nJ_r^2   -
\varepsilon^\dagger(J_r)\right]^s.
\eeq

On the contrary, in the  high $J$ limit, $\moy{J_r}$ is overestimated.
As $J$  increases, the orbital transition state  reaches small values,
which  explains that  the $r_0=0$  radial potential  does not  allow a
correct description of the dissociation process. On the other hand the
harmonic limit tends to  underestimate $\moy{J_r}$ on the whole domain
of $J$  considered in  this study. Therefore  the harmonic  model with
$r_0=0$ could well  give results in quantitative agreement  with MD in
the low $J$ regime, but this would remain fortuitous.

Let $J_{\rm c}$ be the value of the initial angular momentum for which
$\moy{J_r}=J$. For $J<J_{\rm c}$,  evaporation leads to an increase in
the   mean  angular   momentum,  the   so-called   rotational  heating
process. High initial  values of $J$ more likely  result in a decrease
of angular momentum corresponding to rotational cooling. At $E=-29$ LJ
units,  $J_{\rm  c}$  turns  out  to be  approximately  equal  to  2.2
units. As seen  from the previous figures, $J_{\rm  c}$ weakly depends
on the dissociation  potential, but is more sensitive  to the harmonic
approximation to the vibrational density of states.

\begin{figure}[htb]
\vbox to 8cm{
\includegraphics{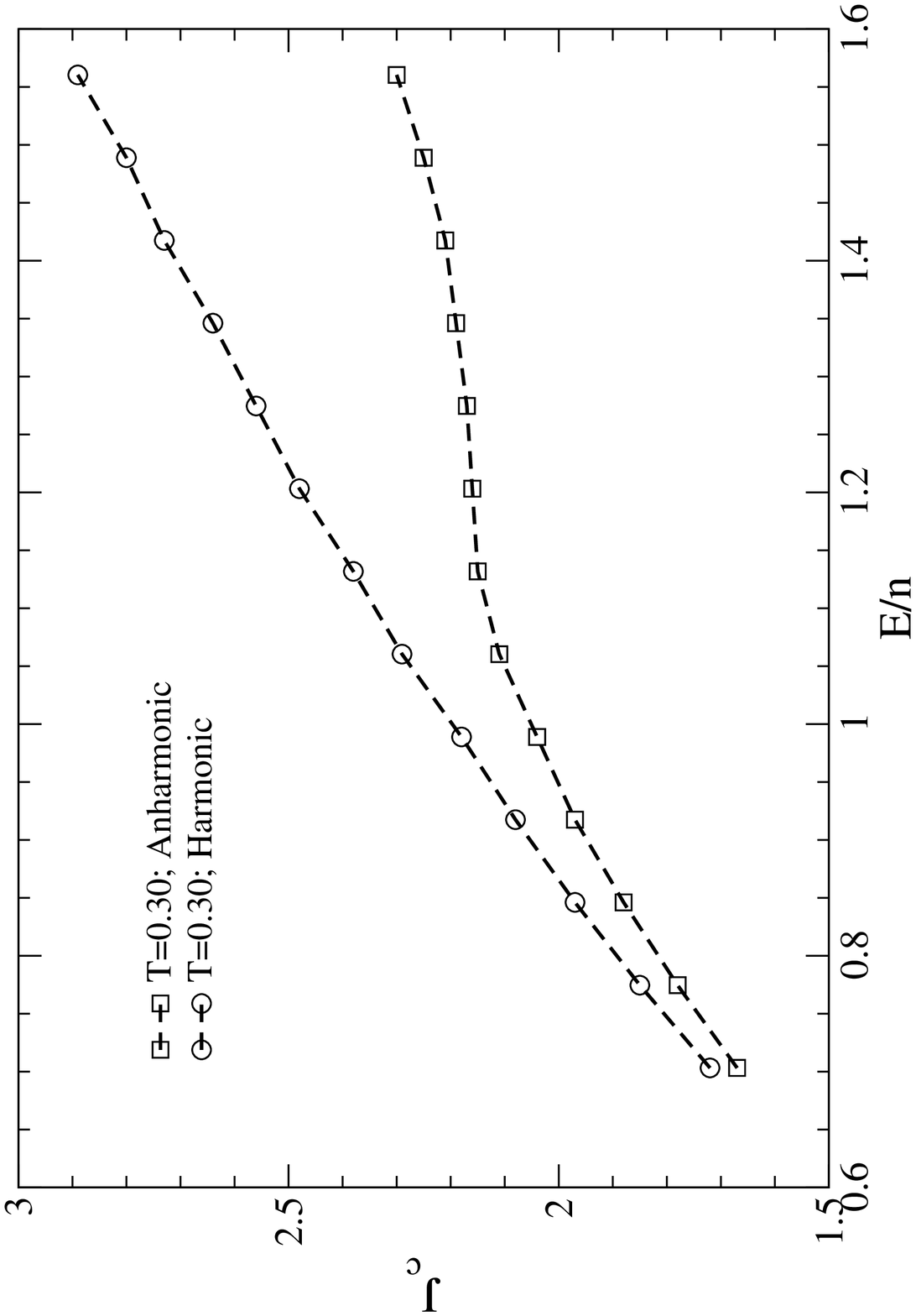}
\vfill}
\caption{$J_{\rm c}$ as a function of $E/n$ for the reaction
LJ$_{14}\longrightarrow$LJ$_{13}+$LJ.}
\label{fig:JcrossingLJ13}
\end{figure}

In Fig.~\ref{fig:JcrossingLJ13} we follow the evolution of $J_{\rm c}$
as a function of the internal energy per atom $E/n$. $J_{\rm c}$ was
calculated with  the numerically exact  radial potential, both  in the
harmonic and anharmonic approximations to  the VDOS. In the low energy
regime, the cluster tends to  be harmonic and both approximations give
similar results.  At higher  energies, especially close to the melting
phase   change   of  LJ$_{13}$   and   beyond,  stronger   differences
appear. Between $E/n=1.1$ and $E/n=1.4$, $J_{\rm c}$ reaches a plateau
similar to the  plateau in the microcanonical temperature  at the same
energy, but only within  the anharmonic description.  The same plateau
must also  be correlated  with the one  in the average  kinetic energy
released, as seen in Refs.~\onlinecite{calvo_2003,wa}, and
\onlinecite{parneix2}. Hence,
large  internal energies  in  the parent  cluster  indeed requires  an
anharmonic treatment of vibration to extract quantitative informations
about the threshold between rotational cooling and heating effects.

\begin{figure}[htb]
\vbox to 8cm{
\includegraphics{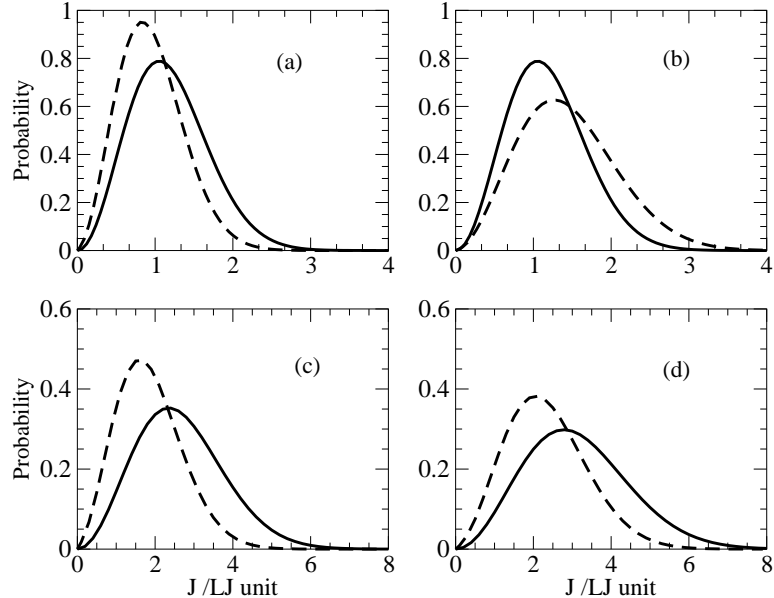}
\vfill}
\caption{Initial and final angular momenta following the evaporation of
LJ$_{14}$ for different physical conditions characterized by their
thermal distributions and the extra excitation energy. (a) $T=0.05$
and $\Delta E=5$; (b) $T=0.05$ and $\Delta E=8$; (c) $T=0.20$ and
$\Delta E=0$; (d) $T=0.40$ and $\Delta E=0$.}
\label{fig:ThermalLJ13}
\end{figure}

Up to now we have only analyzed product angular momentum distributions
for  a given  initial angular  momentum $J$  and an internal
rovibrational energy  $E$.  We will now  consider physically realistic
conditions, where both internal  energy and angular momentum are drawn
from thermal distributions.  The initial angular momentum distribution
$f(J)$ for the spherical cluster  LJ$_{n+1}$ is given by $f(J) \propto
J^2\exp \left( -B_{n+1}J^2/k_BT_{\rm rot}\right)$. The distribution of
the initial  vibrational energy $f(E)$  follows the VDOS, and  we have
allowed for a possible extra amount  of energy $\Delta E$ brought by a
specific     excitation    process     (collision,    photoexcitation,
ionization,...).   Therefore  $f(E)\propto \Omega_{n+1}^{(J)}(E-\Delta
E)\exp\left[ -(E-\Delta  E)/k_B T_{\rm vib}\right]$.  The distribution
of product angular momentum then reads
\beq
P(J_r)\propto \int\!\!\!\!\int J^2 \Omega_{n+1}(E-\Delta E)
\exp\left[ -\frac{E-\Delta E}{k_B T_{\rm vib}} - \frac{B_{n+1}J^2}{k_BT_{\rm rot}}
\right] P(J_r;E,J) dJ dE
\eeq
The initial vibrational and rotational temperatures may be different, because
the extra energy $\Delta E$ is converted into vibrational energy.
In Fig.~\ref{fig:ThermalLJ13} two physical situations have thus been studied.
The first one corresponds to a parent cluster prepared at low temperature
($T=0.05$), for which two possible values of $\Delta E$ have been added,
namely either $\Delta E=4$ or $\Delta E=8$ LJ units.
This situation is typical of a cluster prepared in a supersonic expansion
and photoexcited in the infrared domain. In Fig.~\ref{fig:ThermalLJ13}(a), the
final distribution of $J$ is slighty shifted to lower values.
This is indicative of rotational cooling. In this
case, the mean $E/n$ is almost equal to 0.4. An extrapolation from
Fig.~\ref{fig:JcrossingLJ13} shows that $J_{\rm c}$ will be approximately
equal to 0.9, a value that is smaller than the initial angular momentum.
It explains why cooling occurs under these physical conditions.
On the other hand, in Fig.~\ref{fig:ThermalLJ13}(b), $E/n\approx 0.7$ and
$J_{\rm c}$ is increased to about 1.7. This value is larger than the mean
value of the initial angular momentum, which is consistent with the observed
rotational heating.

We have also  simulated a pure thermal excitation of  the cluster at a
temperature large  enough to induce dissociation on  a reasonable time
scale, that  is accessible to MD. Therefore  the thermal distributions
are directly considered for $E$ and $J$ with $\Delta E=0$, and $T_{\rm
rot}=T_{\rm  vib}$.  We have  considered  two different  temperatures,
$T=0.2$ and $0.4$, and in both cases rotational cooling appears as the
direct consequence  of $\moy{J}  > J_{\rm c}$.   The efficency  of the
cooling  process can  be  analysed by  adjusting  the product  angular
momentum distribution to a Boltzmann law. For example, the thermalized
cluster       initially      at       $T^{(i)}=0.4$       used      in
Fig.~\ref{fig:ThermalLJ13}(d)  yields a  final  rotational temperature
close  to  $T^{(f)}=0.16$.  At  this  initial  temperature,  $E/n$  is
approximately 1, and from Fig.~\ref{fig:JcrossingLJ13} we see that the
harmonic  approximation would  significantly  underestimate rotational
cooling.

Finally  we have  investigated  larger clusters,  within the  harmonic
approximation as well  as the simplest case for  the radial potential,
given by $r_0=0$ and $C_6=4n$  LJ units. The rotational constant $B_n$
has been  taken proportional  to $n^{-5/3}$, with  the proportionality
factor calculated  from the LJ$_{13}$  cluster at $T=0$. The  value of
$E_0$, the  energy difference between LJ$_{n+1}$ and  LJ$_n$, has been
taken to 6.4  for all sizes, a good approximation  in the present size
range of interest.

\begin{figure}[htb]
\vbox to 8cm{
\includegraphics{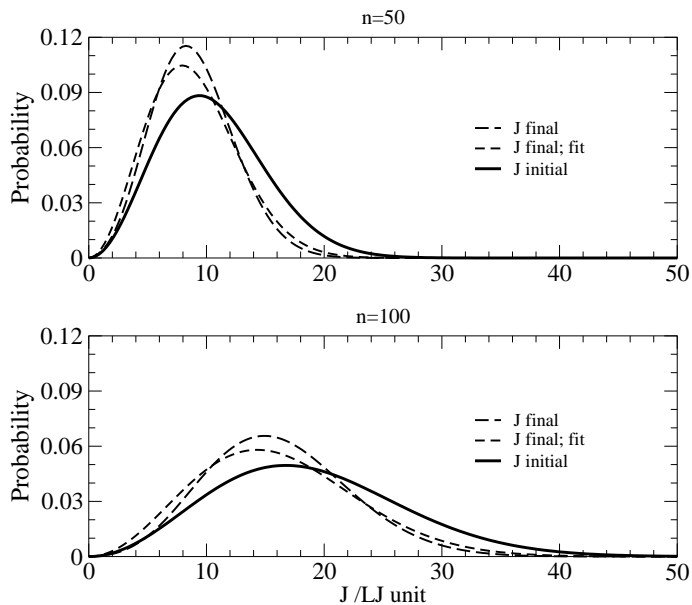}
\vfill}
\caption{Distribution of initial and final angular momenta from
thermal excitations of LJ$_{n+1}$ at $T=0.5$. The best fit to a Boltzmann
distribution is also reported for the product. (a) $n=50$; (b)
$n=100$.}
\label{fig:Size_effect}
\end{figure}

\begin{figure}[htb]
\vbox to 8cm{
\includegraphics{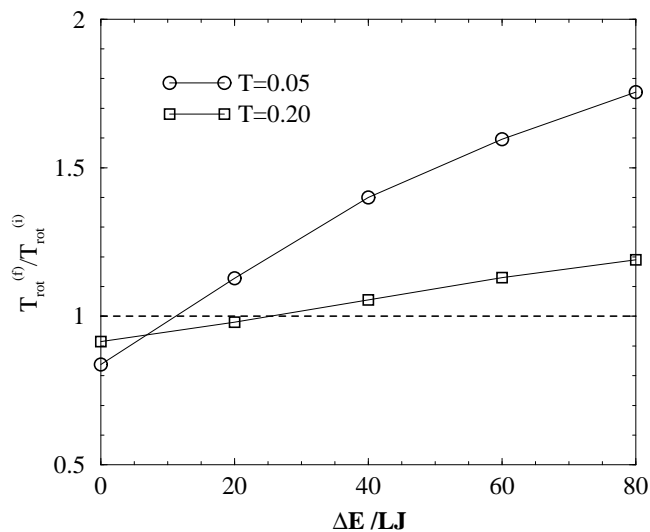}
\vfill}
\caption{Final/initial rotational temperature ratio following
evaporation of LJ$_{51}$, for two thermal distributions at $T=0.05$
and $T=0.20$, as a function of the extra excitation energy.}
\label{fig:comptemp}
\end{figure}

In Fig.~\ref{fig:Size_effect} the angular momentum distribution before
(solid line) and  after (dashed line) evaporation is  shown for $n=50$
and $n=100$  with an initial (vibrational  and rotational) temperature
$T=0.5$. The  adjustment to a  thermal distribution is also  shown. At
this initial temperature, the final rotational temperature is equal to
0.36. It  has to  be noted  that  the final  distribution is  slightly
non-Boltzmann  (more  peaked) with  respect  to  the adjusted  thermal
distribution. Within  the present harmonic approximation,  it is found
that the extent of rotational cooling does not strongly depend on cluster
size.  It is  worth  comparing  the present  approach  to the  results
obtained by Stace on Ar$_n^+$ clusters.\cite{stace91}
The physical conditions chosen by
this author are  those of a cold thermal distribution  of $J$, with an
infinitely  narrow extra  vibrational energy  distribution  located at
$\Delta E$. Because $\Delta E$ is very large in this work (1~eV), many
evaporations  will occur,  and  the width  of  the vibrational  energy
distribution does not really matter.  Hence we can safely take $T_{\rm
vib}=T_{\rm   rot}$.   We    have   calculated   the   ratio   $T_{\rm
rot}^{(f)}/T_{\rm rot}^{(i)}$ between the final and initial rotational
temperatures, for LJ$_{50}$ and  at several excitation energies in the
range  $0\leq \Delta  E\leq  80$ LJ  units.  The value  used by  Stace
corresponds approximately here to $\Delta E\sim 50$--80 units. The results
are  plotted  in Fig.~\ref{fig:comptemp}.  At  small extra  excitation
energies,  $J_{\rm  c}$  is  small  and  evaporation  is  more  likely
accompanied  with rotational  cooling.  On the  other  hand, at  large
values of $\Delta E$, rotational  heating occurs easily. It seems also
pretty obvious that the  smaller the initial temperature, the stronger
the  heating at  a  given excitation  energy.  The transition  between
cooling and  heating occurs near $\Delta E =12$ (for  $T=0.05$) and 
$\Delta E=25$
(for $T=0.20$), and we do not find any evidence for rotational cooling
in LJ$_{50}$ at  excitation energies as large as 50  LJ units. This is
in clear  contradiction with  the results by  Stace,\cite{stace91} who
found heating after  a single evaporation in clusters  smaller than 30
atoms, and  cooling in  40- and 50-atom  clusters at $T\sim  0.05$. At
$T\sim 0.20$,  only the 10-atom cluster  undergoes rotational heating.
This trend  is also  observed in  our work, since  an increase  in the
initial rotational temperature indeed favors cooling.

\section{Discussion and Conclusion}
\label{sec:ccl}

Systematic investigations\cite{wa,ph} have
shown that phase space theory, in the sense of Chesnavich and
Bowers,\cite{cb} performs much better than RRK or Weisskopf-Engelking
models in reproducing the data of reference trajectory simulations.
Weerasinghe and Amar emphasized that the interaction between an atom
and the remaining cluster could not be modelled by a simple $C/r^n$
potential, and that the geometrical extent of the system had to be
taken into account. The detailed studies by Peslherbe and
Hase\cite{ph} on rotating and nonrotating aluminium clusters confirmed
the general agreement between PST and MD, but they noticed some
deviations at large excitation energies. Probably the simple form
chosen for the dissociation potential is not appropriate for these
clusters, for which the interatomic potential (Lennard-Jones pairwise +
Axilrod-Teller three-body component) is also not fully isotropic.

To our knowledge, there has not been any attempt at describing evaporation
in rotating clusters using PST, except in the works by Peslherbe and
Hase.\cite{ph0,ph} One of the interesting results found in the present work
concerns the shape of the distribution of the product angular momentum, which
was seen to have a discontinuity in its derivative for $J=J_r$. The statistics
gathered by Peslherbe and Hase [100 trajectories for a given $(E,J)$\cite{ph}]
are probably not sufficient to confirm this prediction, and we see that
15 sets of 5000 trajectories are necessary to get a hint at this effect.

Here, we have used PST in its general form, assuming that both energy and
angular momentum were conserved during dissociation, calculating transition
state properties at the centrifugal barrier, and without assuming a $C/r^n$
form for the interaction potential between fragments. The main approximation
was to treat the product cluster as a sphere, but this seems appropriate, in
practice, for a large cluster in its liquidlike phase.

The significant differences between the present results and the work
by Stace\cite{stace91} probably comes from the assumption made by this author
that all translational energies released are uniformly distributed. This is a
very strong assumption, which, as a matter of fact, prevents a
rigorous conservation of energy during dissociation. Its consequence
is that the most probable value of $\varepsilon_t$ is $(E-E_0)/2$, much
too large with respect to the simple RRKM estimate given by
$(E-E_0)/s$. Therefore rotational energies are significantly
underestimated, which is precisely observed under the form of
preferential cooling. To confirm these differences between our
work and the results by Stace, we have performed additional
MD simulations of LJ$_{10}$, at the same conditions as above for
$n=50$. In particular, we have carefully looked at the distribution
of final angular momentum of the product LJ$_9$, but we could not
find any signature of bimodality, contrary to the picture
discussed in Ref.~\onlinecite{stace91}. Our PST calculations, within
the harmonic and simple potential approximations, show a
stronger rotational heating than the Stace prediction: for 
$T_{\rm rot}^{(i)}=0.05$ we get $T_{\rm rot}^{(f)}=0.28$ at
an extra excitation energy $\Delta E=25$ units only (in agreement with
MD), instead of $T_{\rm rot}^{(f)}\approx 0.22$ for $\Delta E>50$ units in Stace's
work.\cite{stace91}

Even though the harmonic approximation may be questioned at high energies,
it allowed us to undertake quantitative studies for clusters containing
an arbitrary number of atoms. This required some elementary properties
such as rotational constants or binding energies to be approximated in a
liquid drop fashion, with explicit functions of size. Surely this approach
is too simplistic, as it neglects non-monotonic finite size
effects.\cite{jortner}
The form chosen for the interaction potential may also not be appropriate for
very large sizes, for which the location of the centrifugal barrier becomes
of the same magnitude as the cluster radius itself. Fortunately the
numerical effort
involved in the calculations of the anharmonic DOS and the effective
dissociation potential is comparatively smaller than running a series of
trajectories, especially at low energies where the dissociation rate is
vanishingly small.

The   work  carried   out  in   this   paper  and   in  our   previous
article\cite{calvo_2003} can be straightforwardly extended to the case
of multiple (sequential) evaporations, which arise at large excitation
energies,  and  it  should  be   possible  to  overcome  most  of  the
limitations   that  Stace   had   to  impose.\cite{stace91}   Possible
connections  with the  search for  a liquid-gas  transition  in finite
systems\cite{brech02,schmidt01,poch}        could        then       be
investigated. Another extension to molecular clusters would also be of
great  interest,  as  there  are  presently only  very  few  available
theoretical  results based  on statistical  analyses at  the  level of
phase space theory.

\section*{Acknowledgments}

The authors wish to thank the GDR {\it Agr\'egats, Dynamique et
R\'eactivit\'e} for financial support.


\begin{thebibliography}{99}

\bibitem{brech01} C. Br\'echignac, Ph. Cahuzac, B. Concina, J. Leygnier, B.
Villard, P. Parneix, and Ph. Br\'echignac, Chem. Phys. Lett. {\bf 335}, 34
(2001).

\bibitem{vogel01} M. Vogel, K. Hansen, A. Herlert, and L. Schweikhard,
Phys. Rev. Lett. {\bf 87}, 013401 (2001).

\bibitem{brech02} C. Br\'echignac, Ph. Cahuzac, B. Concina, and J. Leygnier,
Phys. Rev. Lett. {\bf 89}, 203401 (2002).

\bibitem{stace91} A. J. Stace, J. Chem. Phys. {\bf 93}, 6502 (1991).

\bibitem{calvo_2003} F. Calvo, P. Parneix, J. Chem. Phys. {\bf 119}, 256
(2003).

\bibitem{wa} S. Weerasinghe, F.G. Amar, Z. Phys. D: At., Mol. Clusters
{\bf 20}, 167 (1991); J. Chem. Phys. {\bf 98}, 4967 (1993).

\bibitem{ph0} G. H. Peslherbe and W. L. Hase, J. Chem. Phys. {\bf 105}, 7432
(1996).

\bibitem{ph} G. H. Peslherbe and W. L. Hase, J. Phys. Chem. A {\bf 104},
10556 (2000).

\bibitem{pechukas65} P. Pechukas and J. C. Light, J. Chem. Phys. {\bf 42}, 3281
(1965).

\bibitem{light67} J. C. Light, Discuss. Faraday. Soc. {\bf 44}, 14 (1967).

\bibitem{klots71} C. E. Klots, J. Phys. Chem. {\bf 75}, 1526 (1971).

\bibitem{klots76} C. E. Klots, J. Chem. Phys. {\bf 64}, 4269 (1976).

\bibitem{cb1} W. J. Chesnavich and M. T. Bowers, J. Am. Chem. Soc. {\bf 98},
8301 (1976).

\bibitem{cb} W. J. Chesnavich and M. T. Bowers, J. Chem. Phys. {\bf 66},
2306 (1977).

\bibitem{cb2} W. J. Chesnavich and M. T. Bowers, J. Am. Chem. Soc. {\bf 99},
1705 (1977).

\bibitem{schmidt01} M. Schmidt, T. Hippler, J. Donger, W. Kronm\" uller, B.
von Issendorff, H. Haberland, and P. Labastie, Phys. Rev. Lett. {\bf 87},
203402 (2001).

\bibitem{parneix2} P. Parneix, F. G. Amar, and Ph. Br\'echignac, J. Phys.
Chem. {\bf 239}, 121 (1998).

\bibitem{poch} J. Pochodzalla, T. M\"ohlenkamp, T. Rubehn {\em et al.}, Phys.
Rev. Lett. {\bf 75}, 1040 (1995).

\bibitem{mcrot} F. Calvo and P. Labastie, Euro. Phys. J. D {\bf 3}, 229 (1998).

\bibitem{jarrold} M. F. Jarrold, {\em Introduction to statistical reaction
theories}, in Clusters of Atoms and Molecules I, edited by H. Haberland
(Springer, Berlin), 1991.

\bibitem{ptmc} R. H. Swendsen and J.-S. Wang, Phys. Rev. Lett. {\bf 57},
2607 (1986); G. J. Geyer, in {\em Computing Science and Statistics:
Proceedings of the 23rd Symposium on the Interface} (American Statistical
Association, New York, 1991).

\bibitem{histo} A. M. Ferrenberg and R. H. Swendsen, Phys. Rev. Lett. {\bf 61},
2635 (1988).

\bibitem{conversion} One angular momentum LJ unit equals 33.41$\hbar$ for
Argon.

\bibitem{jortner} J. Jortner, Z. Phys. D: At., Mol. Clusters {\bf 24}, 247 (1992).

\end{thebibliography}
\end{document}